\documentclass[12pt]{aipproc}
\layoutstyle{6x9}
\input epsf.tex

\catcode`@=12
\def\aa{{\cal A}}
\begin{document}
\begin{flushright}

\end{flushright}
\vspace{0.5in}
\title{Neutrino Masses and Interactions\\
in a Model with Nambu-Goldstone Bosons
}
\author{E.A. Paschos}{address={Institut f\"ur Physik, Universit\"at Dortmund,
D-44221 Dortmund, Germany},
}

\begin{abstract}\

  A natural scenario for the generation of neutrino masses is the
see-saw mechanism, in which a large right-handed neutrino
mass makes the left-handed neutrinos light.
We review a special case when the Majorana masses originate
from spontaneous breaking of a global U(1)XU(1) symmetry. 
The interactions of the right-handed 
with the left-handed neutrinos at the electorweak scale further
break the global symmetry giving mass to one  
pseudo Nambu-Goldstone boson (pNGB). The pNGB can then generate 
a long-range force. Leptogenesis occurs through decays of 
heavy neutrinos into the light ones and Higgs particles. 
The pNGB can become the acceleron field and the neutrino masses 
vary with the value of the scalar field\footnote{
Invited Talk presented at the Second International Conference on 
High Energy Physics (CICHEP II), January 14-17,2006; Cairo, Egypt.
The talk is a brief preview of the results in reference [11].}.

\end{abstract}
\maketitle

\newpage
\section{Introduction}

The standard model of electroweak theory is succesful in explaining
most of the phenomena we observed till now. The neutrino mass is the
only indication we have that requires physics beyond the standard model. 
It has been established that the neutrino sector must be extended to
accommodate the observed mass-squared difference for the neutrinos
in the atmospheric, solar, as well as in 
the Laboratory experiments. 

The observed neutrino masses are too small to treat them at par with
charged fermions. The most natural explanation for the smallness
of the neutrino masses comes from the see-saw mechanism \cite{seesaw}.
One introduces the right-handed neutrinos in the model that allows
the usual Dirac masses for the neutrinos to be of the order of 
charged lepton masses. Since the right-handed neutrinos are singlets,
it is now possible to allow very heavy Majorana masses $M_1, M_2, ...,$ 
for right-handed neutrinos. This introduces the scale of lepton
number violation in the model and in turn induces tiny Majorana
masses for the left-handed neutrinos. 

When Majorana masses of the right-handed neutrinos are introduced,
they break lepton number explicitly. However, it is more natural to
associate any new scale of the theory with a symmetry  whose breaking
produces neutrino masses. In a forthcoming article we propose [11], that
the Majorana masses of the right-handed neutrinos originate from
the spontaneous breaking of lepton number, which leads to a massless 
Goldstone boson, the Majoron. However, since the residual global
symmetry is not exact, one Goldstone boson picks up small mass 
producing ultra low mass pseudo Nambu-Goldstone bosons (pNGB). 
An early proposal, along these lines, was the axion, which was 
introduced to solve 
the strong CP problem through the Peccei-Quinn symmetry \cite{1,2}. 
The axion is still being searched for as a candidate for dark 
matter \cite{cast}. In strong interactions, it is the breaking of 
symmetries within QCD that generates a large part of the proton's mass, 
which is relatively large compared to the masses of current quarks. 
The result is the appearance of low mass pions -- the pNGB's of the spontaneously broken chiral symmetry. This concept of pNGB was
applied \cite{xx} to the general formalism to a long-range force and phase
transitions \cite{4} occurring at a late epoch of the universe, thus
developing quintessence models contemporaneously with other authors \cite{5}.
Neutrino physics has also found applications of quintessence model, in
which the quintessence or the acceleron are formed as condensates and couple to neutrinos varying their masses \cite{mavans}. This scenario has some interesting consequences \cite{kap,lg}. In this review we point out 
that the pNGB in models of Majorana neutrinos can play the role of the acceleron, introducing an exponential potential, and giving 
rise to a long-range force that may be of of phenomenological importance \cite{new}. Finally, the model generates a lepton asymmetry through the decay of heavy neutrinos to light ones and Higgs particles.

\section{Two Generation PNGB Model}

We demonstrate here the idea behind the pNGB model of Majorana
neutrinos with a two generation example. The theory possesses a
global $U(1)_A \times U(1)_B$ symmetry. The breaking of this
symmetry spontaneously sets a large scale in the theory far above
the scale of electroweak symmetry breaking. There
are two right-handed 
neutrinos $N_1$ and $N_2$, which are singlets under the standard model
gauge groups and transform under the global symmetry as $(1,0)$ and
$(0,1)$ respectively. The global symmetry prevents any Majorana
mass of the right-handed neutrinos. 
We also introduce two singlet scalar fields 
$\Phi_1(x)$ and $\Phi_2(x)$ transforming under the global symmetry
as $(2,0)$ and $(0,2)$ respectively, whose interactions with the
right-handed neutrinos are given by
\begin{equation}
{\cal L}_M = { 1 \over 2} \alpha_1 \bar N_1 N_1^c \Phi_1 +
{ 1 \over 2} \alpha_2 \bar N_2 N_2^c \Phi_2 .
\end{equation}
The vacuum expectation values ($vev$s) of these scalars will give
Majorana masses to the right-handed neutrinos and there will be
two Goldstone bosons. As we shall demonstrate below, the Dirac masses
of the neutrinos at the electroweak scale will break one
combination of the global symmetry softly giving a small mass to
one of the Goldstone bosons, which now becomes a pseudo
Nambu-Goldstone boson (pNGB) of the model. The other Goldstone
boson, the Majoron, remains massless. Since the Majoron 
couples only to singlet fields, the theory is 
consistent with phenomenology of known particles. 

After the fields $\Phi_i, i =1,2$ acquire $vev$s, we can
express them in terms of their $vev$s $\sigma_i$ and decay constants
$f$ (we assume their decay constants to be same, $f_1 \sim f_2 \sim f$)
and the Nambu-Goldstone bosons $\phi_i$ as
\begin{equation}
\alpha _i \Phi_i \to \alpha_i \langle \Phi_i \rangle e^{2 i \phi_i/f_i} 
= \alpha_i \sigma_i e^{2 i \phi_i/f_i} = M_i e^{2 i \phi_i/f_i} .
\label{potn}
\end{equation}
At this stage it is possible to make phase transformations to the
fields $N_i$ that eliminate the  $\phi_i$, implying that 
both Goldstone bosons are massless.
The self interactions of the Goldstone bosons, given by
$$
{\cal L}_\Phi = {1 \over 2} M_\Phi^2 \Phi^\dagger \Phi + {1 \over 4} \lambda
(\Phi^\dagger \Phi)^2 + \partial_\mu \Phi^\dagger \partial^\mu \Phi
= {\sigma^2 \over f^2} \partial_\mu \phi \partial^\mu \phi 
+ f(\Phi^\dagger \Phi)
$$ 
also do not contain any term that can give masses to the Goldstone
bosons.

We now write down all the Yukawa terms involving the neutrinos, including
the Dirac mass terms
\begin{eqnarray}
{\cal L}_{mass} &=& { 1 \over 2} M_1 \bar N_1 N_1^c e^{2 i \phi_1 /f} +
{ 1 \over 2} M_2 \bar N_2 N_2^c e^{2 i \phi_2/f}
+ m e^{i \alpha} \bar N_1 \nu_1 + m \epsilon e^{i \beta} \bar N_1
\nu_2 \nonumber \\
&&+ \lambda m \epsilon^\prime e^{i \gamma} \bar N_2 \nu_1
+ \lambda m e^{ i \xi} \bar N_2 \nu_2 .
\label{dirac}
\end{eqnarray}
We introduced the Dirac mass $m$ and some scaling parameters 
$\lambda, \epsilon, \epsilon^\prime$ in order to write down the Dirac 
mass terms that break the $U(1)_{A-B}$ global symmetry softly.
We included all the phases which contribute to CP violation.
Rephasing of the fields 
\begin{equation}
N_i \to e^{i \phi_2/f} N_i ~~~~{\rm and} ~~~~
\nu_i \to e^{i \phi_2/f} \nu_i
\label{transan1}
\end{equation}
and rephasing of the CP phases, leads to the full mass matrix
\begin{eqnarray}
{\cal L}_\mu &=& {1 \over 2} M_1 \bar N_1 N^c_1 e^{2 i \phi/f}+
{1 \over 2} M_2 \bar N_2 N^c_2 +
m \bar N_1 \nu_1 + m \epsilon e^{i \eta  } \bar N_1
\nu_2 \nonumber \\
&&+ \lambda m \epsilon^\prime e^{i \eta } \bar N_2 \nu_1
+ \lambda m \bar N_2 \nu_2 + H.c. ,
\label{full}
\end{eqnarray}
where $2 \eta = \gamma - \alpha + \beta - \xi$. Thus we finally have
one CP phase $\eta$ and one combination of the fields 
$\phi = \phi_1 - \phi_2$, which becomes the pNGB. The other combination
of  fields $\phi_1 + \phi_2$ correspond to the invisible singlet 
Majoron and remains massless by decoupling from the theory. This conclusion
is independent of the choice of phase transformation.

We shall now explicitly demonstrate how such $U(1)_A \times U(1)_B$ 
global symmetry breaking soft Dirac mass term may appear in the theory after
the electroweak symmetry breaking. If we assign the $U(1)_A \times U(1)_B$
quantum numbers 
$$\ell_1 \equiv \pmatrix{ \nu_e \cr e^-} \equiv (1,0)
~~~~{\rm and} ~~~~\ell_2 \equiv \pmatrix{ \nu_\mu \cr \mu^-}
\equiv (0,1), $$ then
the usual Higgs doublet $H$ with the assignment $(0,0)$ (for this to
give masses to the charged fermions) can give only the diagonal Dirac 
mass terms. We need two more Higgs doublets $H_1 \equiv (-1,+1)$ and
$H_2 \equiv (+1,-1)$ in order to generate the complete Dirac mass we discussed above. The mass term of the Lagrangian now becomes
\begin{eqnarray}
{\cal L}_{mass} &=& { 1 \over 2} M_1 \bar N_1 N_1^c e^{2 i \phi_1 /f} +
{ 1 \over 2} M_2 \bar N_2 N_2^c e^{2 i \phi_2/f}
+ f_{11} \bar N_1 \nu_1 H + f_{12} \bar N_1
\nu_2 H_1\nonumber \\
&&+f_{21} \bar N_2 \nu_1 H_2 + f_{22} \bar N_2 \nu_2 H .
\label{dirac1}
\end{eqnarray}
These dimension-4 terms do not break the global symmetry. However,
after the electroweak symmetry breaking, when all the doublets $H$
and $H_i$ acquire $vev$s, the global symmetry is broken. The
part of the Lagrangian with all neutrino mass terms is:
\begin{eqnarray}
-{\cal L}_{mass} &=& { 1 \over 2} M_1 \bar N_1 N_1^c e^{2 i \phi_1 /f} +
{ 1 \over 2} M_2 \bar N_2 N_2^c e^{2 i \phi_2/f}
+ m e^{i \alpha} \bar N_1 \nu_1 + m \epsilon e^{i \beta} \bar N_1
\nu_2 \nonumber \\
&&+ \lambda m \epsilon^\prime e^{i \gamma} \bar N_2 \nu_1
+ \lambda m e^{ i \xi} \bar N_2 \nu_2 .
\label{dirac}
\end{eqnarray}

\begin{figure}[!h]
\epsfxsize10cm\epsffile{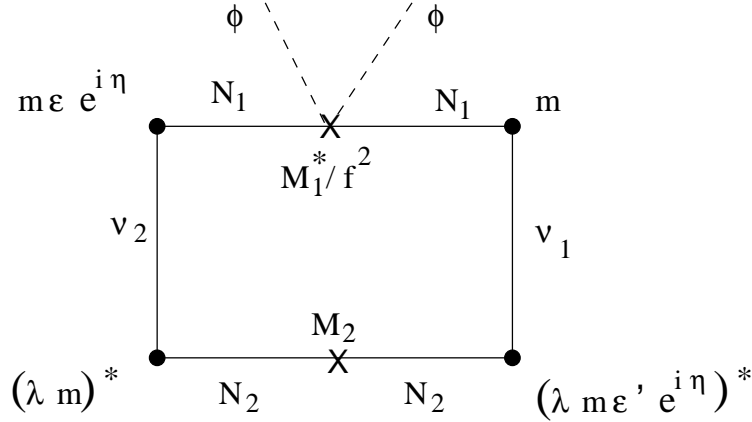} 
\caption{Loop diagram for the effective potential
}
\label{figure1}
\end{figure}
After rephasing of the fields it reduces to equation (5).

To find out the mass of the pNGB, we consider the effective potential
generated by the interactions of the scalar field through the mass terms.
The Colemann-Weinberg potential for $\phi$ is computed through the
leading loop in Fig. 1. It has the remarkable property that the symmetry
structure of the theory makes the loop finite. The reason is that the $\phi$
field could be eliminated if any of the vertices is set to zero. 
This diagram is also invariant under any rephasing of the fields,
which can be confirmed by observing that the most general phase 
transformation of the neutrinos,
\begin{equation}
N_i \to e^{i p_i} N_i ~~~~ {\rm and} ~~~~ \nu_i \to e^{i q_i} ,
\end{equation}
transforms 
\begin{eqnarray}
m \to e^{i (q_1 - p_1)} ~m; &~~~& 
m \epsilon \to e^{i (q_2 - p_1)} \nonumber \\ 
m \lambda \epsilon^\prime \to e^{i (q_1 - p_2)} ~m 
\lambda \epsilon^\prime;  &~~~& \lambda m \to
e^{i (q_2 - p_2)} ~\lambda m ; \nonumber \\
M_1 \to e^{-i 2 p_1 } ~M_1;  &~~~& 
M_2 \to e^{-i 2 p_2 } ~M_2;
\end{eqnarray}
so that 
$$ (m) \cdot (M_1^*/f^2) \cdot (m ~\epsilon ~e^{i \eta}) \cdot
(\lambda ~m)^* \cdot (M_2) \cdot (\lambda ~m ~\epsilon^\prime ~
e^{i \eta} )^* $$
and hence the 
diagram remains invariant. 

The explicit calculation gives 
\begin{equation}
V_{eff} (\phi^2) = - {m^4 \lambda^2 \epsilon \epsilon^\prime \over
4 \pi^2} {M_1 M_2 \log \left( {M_1^2 \over M_2^2} \right) 
\over M_1^2 - M_2^2 } \cos \left({ 2 \phi \over f} \right),
\end{equation}
which has the minima at $\phi = 0, \pi f, 2 \pi f, \cdot \cdot \cdot$. 
Expanding around one of the minima, we can write down 
a mass term 
\begin{equation}
V_{eff} (\phi) = {m^4 \lambda^2 \epsilon \epsilon^\prime \over
2 \pi^2} {M_1 M_2 \log \left( {M_1^2 \over M_2^2} \right) 
\over M_1^2 - M_2^2 } {\phi^2 \over f^2} 
\end{equation}
Thus the mass of the $\phi$ field is now
\begin{equation}
m_\phi = {m^2 \lambda \sqrt{\epsilon \epsilon^\prime} \over
\pi f} {M_1 M_2 \log \left( {M_1^2 \over M_2^2} \right) 
\over M_1^2 - M_2^2 } .
\end{equation}
The symmetry at the scale of $f \sim M_i$ protects
the mass of the pseudo Nambu-Goldstone boson, so an explicit soft breaking
of the symmetry at the scale $m$ can generate a mass of the order of
$m^2/f$. This light pNGB ($\aa = i \phi$) 
can generate a long-range force and also
become the acceleron field to explain the smallness of dark energy.

\section{Neutrino Masses}

We consider next the structure of neutrino masses in this model.
After the global symmetries are broken by the $vev$s of the
scalars, the model reduces to the usual see-saw models \cite{seesaw}, except
for the pNGB which generates a new long-range force because of its
small mass. The interactions required for the usual leptogenesis \cite{lepto,lepto1,lepto2} through the decays of right-handed 
neutrino are also present and will be discussed in the next section.

We shall first consider the time evolution of the light neutrino
states without including the effects of the pNGB, which is
determined by the matrix
\begin{equation}
- {\cal L}_{eff} = m_{ij}^T M_i^{-1} m_{ij}
~ \nu_i \nu_j = {m^2 \over M}\left[ {\pmatrix{\nu_1 & \nu_2} \atop }
\pmatrix{ 1 + (\lambda \epsilon^\prime)^2
e^{2 i \eta } & e^{ i \eta} ( \epsilon + \lambda \epsilon^\prime) \cr
e^{ i \eta} ( \epsilon + \lambda \epsilon^\prime) & \lambda^2 + \epsilon^2
e^{2 i \eta }  } \pmatrix{\nu_1 \cr \nu_2} \right]. 
\end{equation}
This gives a mixing angle
\begin{equation}
\tan 2 \theta = 2 {\epsilon + \lambda \epsilon^\prime \over
(1 - \lambda^2 ) e{- i \eta} + (\lambda^2 {\epsilon^\prime}^2 -
\epsilon^2)e{ i \eta}}
\end{equation}
which is large for $\epsilon, \epsilon^\prime \ll 1$
and $\lambda \approx 1$. This mass matrix can be diagonalized to
\begin{equation}
M^{diag} = \pmatrix{m_1 e^{ i \theta_1} & 0 \cr 0 &m_2 e^{ i \theta_2}}
\end{equation}
with
\[
m_{1,2} = {m^2 \over M} [1 \pm (\epsilon + K) + 
O( \epsilon^2)], ~~~{\rm and} ~~~
\tan \theta_{1,2} = {1 \pm (\epsilon + K) \cos \eta \over
\pm (\epsilon + K) \sin \eta },
\]
where $K = \lambda \epsilon^\prime$. This part of the discussion is
the same as in any other model with a see-saw mechanism. It can account 
for the large mixing between two generations. 

We shall now consider the effect of the light scalar field, the pNGB
in the model. We obtain the interactions of the light neutrinos with
the scalar field by keeping only the terms linear in pNGB, given by
\begin{equation}
{\cal L} = {m^2 \over 2 M} ~ \overline{\Psi^c} \left\{ \pmatrix{
m_1 &0\cr 0 & m_2} - {\phi \over f} e^{i \eta} \pmatrix{
(\epsilon - K) e^{-i \theta_1} & 0 \cr 0 &(K - \epsilon) e^{-i \theta_2}}
\right\} \Psi + H.C.
\end{equation}
The long range force introduced by the pNGB in this
model will have direct consequences in neutrino oscillation experiments
\cite{bhm05}. The extension of the model to three generations and the implications of the new force will be discussed elsewhere [11].

\section{Lepton Asymmetry of the Universe}

The direct connection between the neutrino masses and the baryon
asymmetry of the universe is now well established. In our case,
 lepton number violating couplings were introduced in order to 
give masses to the 
neutrinos. These couplings generate a lepton asymmetry \cite{lepto,lepto1,lepto2}, 
which will be converted to a baryon number asymmetry as the
universe expands and approaches the electroweak phase transition. 

 In general, the generation of the lepton asymmetry of the universe
requires three ingredients: 
\begin{enumerate}
\item Lepton number violation, which is also the
source of neutrino masses; 
\item CP violation, which comes from the interference of tree level
and one loop diagrams
\item Departure of the lepton number violating interactions from
equilibrium
\end{enumerate}
The decays of the right-handed neutrinos
violate lepton number and the Yukawa couplings contain a phase $\eta$
which can give  CP-even and CP-odd amplitudes. If the right-handed 
neutrino mass can now satisfy the out-of-equilibrium condition, this 
model will be able to generate the required baryon asymmetry 
of the universe.

In our model lepton number is violated at tree-level in decays of the right-handed neutrinos
\begin{eqnarray} 
N_i &\to& \ell_j + H_a^\dagger \nonumber \\
&\to& \ell_j^c + H_a  \nonumber
\end{eqnarray}
The one loop diagrams, like vertex corrections and self energies \cite{lepto,lepto1}
will contain CP-even and -odd amplitudes. The structure of the
Lagrangian in eqs (5) and (6) has loops only when the Higgs particles 
mix with each other. Let us denote by Vo1 the mixing between 
$H_0$ and $H_1$ and similarly by $V_{02}$ the mixing between 
$H_0$ amd $H_2$. Such mixings are generated by the quartic couplings 
of the potential when the scalar acquire vevs. Here we do not 
restrict ourselves to a specific form of the potential and represent 
the generic mixings by $V_{0i}$. The interference between tree 
and vertex diagrams gives the asymmetry
\begin{eqnarray}
\delta &=&  { \Gamma ( N_1 \to \ell H^\dagger) -
\Gamma ( N_1 \to \ell^c H) \over \Gamma ( N_1 \to \ell H^\dagger) +
\Gamma ( N_1 \to \ell^c H)} \nonumber \\[.05in]
&=&
\frac3{16\pi}\frac{M_1}{M_2}\frac{m^2\lambda^2}{v_1^2+v_0^2\epsilon^2}
\left[
|V_{02}|^2\frac{v_1^2}{v_2^2}\epsilon^{'2}-
|V_{01}|^2\epsilon^{2}
\right]\sin 2\eta
\end{eqnarray}
where $v_a = \langle H_a \rangle, a = 0,1,2$ and we assumed
$M_2 >> M_1$, so that the asymmetry is generated by the decays
of $N_1$. 
In the context of leptogenesis the lightest $N_1$ is naturally out of 
thermal equilibrium when it decays. The dilusion factor depends on 
the relative magnitude of 
\begin{equation}
\Gamma_{N_1} = {|f_{1 i}|^2 \over 16 \pi} M_1 
\end{equation}
relative to the Hubble constant
\begin{equation}
 H(M_1) =
1.7 \sqrt{g_*} {T^2 \over M_P} \hskip .5in {\rm at}~T = M_1 .
\end{equation}
When the ratio $\kappa=\Gamma/H(M_1)$ is small the lepton excess is large. 
On the other hand, for large $\kappa$ the time of expansion is large 
relative to the lifetime of $N_1$ so that many decays and recombinations 
can take place in the cosmological scale  $1/H$ giving
a smaller excess of leptons. Finally, the sphaleron interactions 
convert a fraction of the produced ${(B-L)}$ asymmetry to a baryon 
asymmetry 
\begin{equation}
{n_B \over s} = {24 + 4 n_H \over 66 + 13 n_H} {n_{B-L} \over s}.
\end{equation}
This conversion takes place from the time of leptogenesis down to the time of the electroweak phase transition. Here 
$n_H$ is the number of Higgs doublets in our
model. In summary, it is evident that the model can generate 
the required amount of baryon asymmetry of the universe.

\section{Origin of Dark Energy}

One of the most challenging questions at present is why the cosmological
constant is very small but nonvanishing. If the dark energy were due to
the presence of a cosmological constant, then one expects
$$ \rho_{DE} = \rho_{vac} = E_0^4$$
where $E_0$ is the energy associated with a particle or a field. 
The observed cosmological constant coresponds to an  
$E_0 \sim 2 \times 10^{-3}$ eV, which is very small for the scales 
of most elementary particles. It is of the order of
neutrino masses and it has recently been proposed
that the dark energy tracks the neutrino density \cite{mavans}.
If the neutrino masses vary as functions of a light
scalar field (called the acceleron) the dark energy density 
may track the neutrino masses. This would then explain why the 
scale of cosmological constant is the same as the scale of 
neutrino masses. 
   In the present model, the pNGB introduces a spatial dependence
into the Majorana mass M1
\begin{eqnarray}
{\cal L}_\mu &=& {1 \over 2} M_1(\aa) \bar{N_1^c} N_1 +
{1 \over 2} M_2 \bar{N_2^c} N_2 +
m \bar N_1 \nu_1 + m \epsilon \bar N_1
\nu_2 \nonumber \\&&
+ \lambda m \epsilon^\prime \bar N_2 \nu_1
+ \lambda m \bar N_2 \nu_2 + H.c.
\label{full1}
\end{eqnarray}
The first term $M_1(\aa) $ has an exponential functional dependence
on the scalar field.
The effective neutrino mass then varies explicitly with
the acceleron field $m_\nu (\aa) = m^T M^{-1} (\aa) m$, 
as required by models of mass varying neutrinos \cite{mavans}.

\section{Summary}

   A large Majorana scale is a necessary ingredient of the see-saw 
mechanism. At this very high energy there may exist global symmetries 
which are broken in order to create the Majorana masses. 
Remnants of the global 
symmetry survive in the low energy interactions. 
In a two generation model developed with my
collaborators Hill, Mocioiu and Sarkar \cite{new}, the breaking 
of the symmetry produces two Nambu-Goldstone bosons.
 The introduction of Dirac masses 
at the electroweak scale further breaks the symmetry softly giving a 
small and finite mass to one of the Nambu-Goldstone bosons and 
leaves the other scalar massless. The exchange of the scalar 
particle introduces a new long-range force between neutrinos.
 
    Finally, the model has additional attractive properties. It describes
correctly the masses and mixings of the light neutrinos. The decays 
of the heavy neutrinos generate a lepton asymmetry consistent 
with leptogenesis. The neutrino masses have an explicit dependence on the 
scalar field, which may bring density effects with the pNGB playing 
the role of the "acceleron".

\section*{Acknowledgement}
    I wish to thank my collaborators Drs. C.T. Hill, I. Mocioiu and
U. Sarkar for a pleasant and fruitful collaboration. An extensive article mentioned in ref.  [11] is in preparation. I also thank
Dr. Y-F. Zhou and Mr. A. Karavtsev for working with me on extensions
of the model. The financial support of BMBF, Bonn under contract
05HT 4PEA/9 is gratefully ackonwledged. Finally, I thank the DFG 
for a travelling grant to participate at the conference.

\bibliographystyle{unsrt}

\end{document}